\documentclass[twoside]{ilcws10}
\usepackage[latin1]{inputenc}
\usepackage[dvips]{graphicx, epsfig, color}
\usepackage{wrapfig, rotating}
\usepackage{amssymb, amsmath, array}

\def\Slash#1{{\ooalign{\hfil$#1$\hfil\crcr\hfil$/$\hfil}}}

\begin{document}
\title{Observing the Coupling between \\ Dark Matter and Higgs Boson at the ILC}

\author{Shigeki Matsumoto$^1$, Keisuke Fujii$^2$, Takahiro Honda$^3$, Shinya Kanemura$^1$, \\
        Takehiro Nabeshima$^1$, Nobuchika Okada$^4$, Yosuke Takubo$^3$, and Hitoshi Yamamoto$^3$
\vspace{.3cm}\\
1- Department of Physics, University of Toyama, \\
Toyama 930-8555, Japan 
\vspace{.1cm}\\
2- High Energy Accelerator Research Organization (KEK), \\
Tsukuba 305-0801, Japan
\vspace{.1cm}\\
3- Department of Physics, Tohoku University, \\
Sendai 980-8578, Japan
\vspace{.1cm}\\
4- Department of Physics and Astronomy, University of Alabama, \\
Tuscaloosa, AL 35487, USA\\
}

\maketitle

\begin{abstract}
One of the main purposes on physics at the International Linear Collider (ILC) is to study the property of dark matter such as its mass, spin, and interactions between the dark matter and particles in the standard model. Among those, the interaction between dark matter and higgs boson is important, because, if the dark matter is a weakly interacting massive particle, the existence of the dark matter is expected to relate to physics of electroweak symmetry breaking. In addition, the interaction plays an important role at direct detection experiments of dark matter. In this report, we discuss how accurately the coupling constant of the interaction can be measured at the ILC with the center of mass energy less than 500 GeV.
\end{abstract}

\section{Introduction}

The dark matter, which accounts for about 23\% of the energy density in the present Universe~\cite{Komatsu:2008hk}, has been a great mystery in astrophysics, cosmology, and particle physics. While various theoretically-motivated candidates have been discussed for dark matter particle, its detailed nature is still un-revealed. One of the main purposes at collider experiments is to reveal the property of the dark matter such as its mass, spin, gauge charges, and interactions between the dark matter and particles in the standard model (SM). It is, for instance, expected that the mass of the dark matter can be measured at the Large Hadron Collider (LHC) at various candidates for dark matter. On the other hand, the International Liner Collider (ILC) enables us not only to measure the mass very accurately, but also to determine the spin of the dark matter~\cite{DM at ILC}. It may be even possible to measure the coupling constant of each interaction between dark matter and particles in the SM at the ILC.

Among the interactions, the most important one is the coupling between dark matter and higgs boson. This is because, if the dark matter is a weakly interacting massive particle (WIMP), its existence is expected to deeply connect with physics of the electroweak symmetry breaking, to be more precisely, the mechanism to solve the hierarchy problem concerning quadratically divergent corrections to the higgs mass term. In addition, the interaction is also important for direct detection measurements of dark matter such as CDMS~\cite{CDMS} and XENON~\cite{XENON} experiments, since the scattering of dark matter with ordinary matter occurs mainly due to the exchange of the higgs boson between dark matter and nucleon.

It is, therefore, very important to estimate how accurately the coupling constant can be measured at the ILC. We consider the higgs portal scenario for dark matter as a typical example describing interaction between dark matter and higgs boson. With the use of the scenario, we clarify the parameter region on the (dark matter mass, coupling constant)-plane which can be measured at the ILC with the center of mass energy less than 500 GeV.

\section{Setup}

In the higgs portal scenario that we use in this report, the global $Z_2$ symmetry (parity) is postulated in order to guarantee the stability of dark matter, where the WIMP dark matter has odd charge while particles in the SM have even one. We consider three cases for the spin of the dark matter; the scalar dark matter $\phi$, the fermion dark matter $\chi$, and the vector dark matter $V_\mu$. In all cases, the dark matter is assumed to be an identical particle for simplicity, so that these are described by real Klein-Gordon, Majorana, and real Proca fields, respectively. The Lagrangian in the scenario is written as 
\begin{eqnarray}
 {\cal L}_S
 &=&
 {\cal L}_{\rm SM} + \frac{1}{2} \left(\partial \phi\right)^2 - \frac{M_S^2 }{2} \phi^2
 - \frac{c_S}{2}|H|^2 \phi^2 - \frac{d_S}{4!} \phi^4,
 \label{Lagrangian S}
 \\
 {\cal L}_F
 &=&
 {\cal L}_{\rm SM} + \frac{1}{2} \bar{\chi} \left(i\Slash{\partial} - M_F\right) \chi
 - \frac{c_F}{2\Lambda} |H|^2 \bar\chi \chi
 - \frac{d_F}{2\Lambda} \bar{\chi}\sigma^{\mu\nu}\chi B_{\mu\nu},
 \label{Lagrangian F}
 \\
 {\cal L}_V
 &=&
 {\cal L}_{\rm SM} - \frac{1}{4} V^{\mu\nu} V_{\mu \nu} + \frac{M_V^2 }{2} V_\mu V^\mu
 + \frac{c_V}{2} |H|^2 V_\mu V^\mu - \frac{d_V}{4!} (V_\mu V^\mu)^2,
 \label{Lagrangian V}
\end{eqnarray}
where $V_{\mu\nu} = \partial_\mu V_\nu - \partial_\nu V_\mu$, $B_{\mu\nu}$ is the field strength tensor of the hypercharge gauge boson, and ${\cal L}_{\rm SM}$ is the Lagrangian of the SM with $H$ being the higgs boson. The last terms in RHS in Eqs.(\ref{Lagrangian S}) and (\ref{Lagrangian V}) proportional to coefficients $d_S$ and $d_V$ represent self-interactions of the WIMP dark matter, which are not relevant for the following discussion. On the other hand, the last term in RHS in Eq.(\ref{Lagrangian F}) proportional to the coefficient $d_F$ is the interaction between WIMP dark matter and the hypercharge gauge boson, however this term is most likely obtained by 1-loop diagrams  of new physics dynamics at higher energy scale, since the dark matter particle carries no hypercharge. The term therefore can be ignored in comparison with the term proportional to $c_F$ which can be obtained by tree-level diagrams. After the electroweak symmetry breaking, masses of the dark matters are given by
\begin{eqnarray}
 m_S^2 &=& M_S^2 + c_Sv^2/2, \\
 m_F   &=& M_F   + c_Fv^2/(2\Lambda), \\
 m_V^2 &=& M_V^2 + c_Vv^2/2,
\end{eqnarray}
where the vacuum expectation value of $H$ is $\langle H \rangle = (0,v)^T/\sqrt{2}$ with $v$ being $v \simeq 246$ GeV. Model parameters relevant to following discussions are, thus, $m_{\rm DM}$ and $c_{\rm DM}$.

There are some examples of new physics models realizing the higgs portal scenario at low energies~\cite{higgs portal}. The scenario with the scalar higgs portal dark matter is discussed frequently. R-parity invariant supersymmetric standard models with the Bino-like lightest super particle can correspond to the fermion higgs portal dark matter scenario when the other super-partners are heavy enough~\cite{SUSY}. The vector dark matter can be, for example, realized in such as the littlest higgs model with T-parity if the breaking scale is very high~\cite{LHT}.

\section{Signals at the ILC}

We are now in position to investigate the signal of the WIMP dark matter at the ILC experiment. The main purpose of this report is to clarify the parameter region where the signal can be detected. We first consider the case in which the mass of the dark matter is less than a half of the higgs boson mass ($m_h$). In this case, the dark matter particles can be produced through the decay of the higgs boson. Then, we consider the other case where the mass of the dark matter particle is heavier than $m_h$/2.

\subsection{The case $m_{\rm DM} < m_h/2$}

In this case, the coupling of the dark matter particle with the higgs boson can cause a significant change in the branching ratio of the higgs boson while the production process of the higgs boson at the ILC remains the same. When the mass of the higgs boson is not heavy ($m_h \lesssim 150$ GeV), its partial decay widths into quarks and leptons are suppressed due to small Yukawa couplings. As a result, the branching ratio into dark matter particles can be almost 100\% unless the interaction between the dark matter and the higgs boson is too weak. In this case, the most of the higgs bosons produced at the ILC decay invisibly. There are some studies on the invisible decay of the higgs boson at the ILC. The most significant process for investigating such a higgs boson at the center of mass energy less than 500 GeV is found to be the higgs-strahlung process, $e^+ e^- \rightarrow Z H$. In Ref.~\cite{ Schumacher:2003ss}, it has been shown that, using the 500 fb$^{-1}$ data with the center of mass energy of 350 GeV, the invisible decay of the higgs boson can be detected at the 95\% C.L. when the branching ratio of the decay is larger than 0.95\%, where the higgs mass is fixed to be 120 GeV. With the use of the analysis, we plot the experimental sensitivity to detect the signal in Fig.\ref{fig:results}. The sensitivity is shown as blue lines with $m_{\rm DM} \leq m_h/2$, where the signal can be observed at the ILC in the regions above these lines. For comparison, we also plot the sensitivity of the LHC to detect the invisible decay at the 95\% C.L. as green lines, where 30 fb$^{-1}$ data with the center of mass energy of 14 TeV is assumed in the analysis~\cite{Inv H LHC}.

\subsection{The case $m_{\rm DM} \geq m_h/2$}

\begin{figure}[t]
\centerline{\includegraphics[width=\columnwidth]{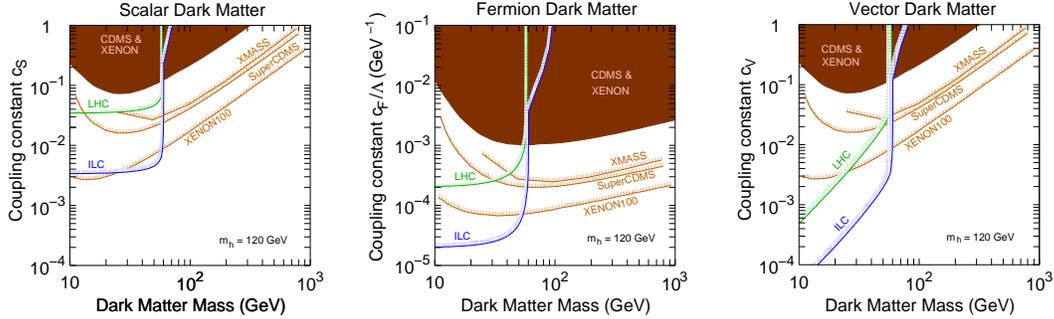}}
\caption{\small Sensitivities to detect the dark matter signal at the LHC and ILC. Constraints and expected sensitivities on direct detection experiments for dark matter are also shown.}
\label{fig:results}
\end{figure}

In this case, the WIMP dark matter cannot be produced from the decay of the higgs boson. We consider, however, the higgs-strahlung process again. The signal we are focusing on in this report is from the process $e^+ e^- \rightarrow h^*Z \rightarrow 2{\rm DM} + q\bar{q}$ with $h^*$ being a virtual higgs boson. The signal is characterized by two energetic quark jets with large missing energy. On the other hand, there are several backgrounds against the signal; $e^+ e^- \rightarrow$ $W^+ W^-$, $Z Z$, $\nu \nu Z$, $e^+ e^- Z$, and $e^\pm \nu W^\mp$, where all of these mimic the signal at the detector. In order to reduce these backgrounds, we apply kinematical cuts on the $Z$ boson reconstructed from two tagging jets, and perform the likelihood analysis~\cite{Ours}. From the analysis of these backgrounds, it turns out that, at the ILC with the center of mass energy of 300 GeV and the integrated luminosity of 2 ab$^{-1}$, the signal will be detected at 95\% C.L. when the cross section of the signal exceeds 0.7-0.8 fb after applying these kinematical cuts. In the analysis, the higgs mass is fixed to be 120 GeV, while the dark matter mass is varied from 60 to 90 GeV. With this analysis, we have estimated the experimental sensitivity to detect the signal at the ILC when $m_{\rm DM} \geq m_h/2$. The result is, again, shown in Fig.\ref{fig:results} as blue lines for $m_{\rm DM} \geq m_h/2$, where, with an integrated luminosity of 2 ab$^{-1}$ and the center of mass energy of 300 GeV, the signal can be observed at 95\% C.L. in the regions above these lines.

\section{Summary and Discussions}

We have investigated signals of the WIMP dark matter at the ILC with, especially, focusing on the coupling between dark matter and higgs boson. We have used the higgs portal scenario as a typical example describing the coupling, and clarify the parameter region which can be detected at the collider with the center of mass energy less than 500 GeV. All results are summarized in Fig.\ref{fig:results}. When the mass of the dark matter is less than $m_h/2$, most of parameter regions can be covered at the ILC using the invisible decay of the higgs boson. On the other hand, if $m_{\rm DM} \geq m_h/2$, the observation of the coupling will be challenging, but still be possible if the coupling constant is large enough. When we use more sophisticated analysis or accumulate more data, the sensitivity may be upgraded.

It is also interesting to compare these results with those of direct detection experiments of dark matter. The result from CDMS II~\cite{CDMSII} and the new data from the XENON 100 experiment~\cite{ Aprile:2010um} give the most severe constraint on the scattering cross section between dark matter particle and nucleon. The result of the constraint is also shown in Fig.\ref{fig:results}, where the regions in brown are excluded by the experiments at 90\% C.L.. In Fig.\ref{fig:results}, we also depict experimental sensitivities to detect the signal of the dark matter in near future experiments, XMASS~\cite{Abe:2008zzc}, SuperCDMS~\cite{Brink:2005ej}, and Xenon100~\cite{Aprile:2009yh}. The sensitivities are shown as light brown lines, where the signal can be discovered in the regions above these lines at 90\% C.L.. Interestingly, the signal of the WIMP dark matter can be obtained in both direct detection measurement and ILC experiment when $m_{\rm DM} \leq m_h/2$, which allow us to perform a non-trivial check for the coupling between dark matter and higgs boson. In the case of $m_{\rm DM} \geq m_h/2$, unfortunately, parameter regions which can be detected at the ILC are almost excluded by current direct detection experiments of dark matter, however more sophisticated analysis for the signal of the ILC may enhance the ratio of the signal to background~\cite{Ours}.

\section{Acknowledgments}

The authors would like to thank all the members of the ILC physics subgroup~\cite{ILC physics subgroup} for useful discussions. This study is supported, in part, by the Grant-in-Aid for Science Research, Ministry of Education, Culture, Sports, Science and Technology, Japan (No. 22244031, and Nos. 21740174 and 22244021 for SM).



\end{document}